# Non-Suicidal Self-Injury Online Posts: Implications for Mental Health Professionals

Mandy Greaves https://orcid.org/0000-0003-0614-5916
Cass Dykeman https://orcid.org/0000-0001-7708-1409

## Abstract

While non-suicidal self-injury (NSSI) is not a new phenomenon, there is still a limited understanding of the behavior, intent behind the behavior and what individuals themselves say about their behavior. This study collected pro-NSSI public blog posts from Reddit and analyzed the content linguistically using LIWC software, in order to examine the use of NSSI specific words, linguistic and psychological linguistic properties. The results inform current mental health practices by dispelling myths and providing insight into the inner world of people who engage in NSSI. The linguistic properties found in the analysis reflected the predicted results; authors of pro-NSSI posts used first-person singular pronouns extensively, indicating high levels of mental health distress and isolation. This study does not encourage or endorse mental health professionals to look into their client's social media use. Rather, this study has generated knowledge and demonstrates how language can become a key factor when assessing a person's NSSI use.

*Keywords:* NSSI, LIWC, linguistic analysis, mental health, self-injury

## Introduction

Non-suicidal self-injury (NSSI) is not a new phenomenon. In his Lectures on Ethics and Metaphysics of Morals, Immanuel Kant (1804) concludes that "We must reverence humanity in our own person" and argues that those "who harm themselves harm all of humanity" (as cited in Portmann, 2004, p. 19). It was not until the 1960's that mental health professionals began to focus on wrist-cutting and classified this behavior as a specific disorder that could be distinguished from both cutting behavior and suicidal behavior. In their study of wrist-cutting syndrome, Rosenthal et al. found in their 1972 study that young women who performed self-harming behaviors described feeling a sense of numbness, dissociation and emptiness leading up to the cutting behavior. Favazza (1998) argued that self-harm is distinguished from suicidal acts by the individual's intentions; a person who is suicidal intends to end all feelings, but a person who self-harms is attempting to cope with negative feelings. While NSSI as a separate, distinctive phenomenon is now well-documented, little else is known about this behavior in adolescents or emerging adults.

Social media use amongst adolescents and emerging adults is at an all-time high. The Pew Research Center survey found that younger Americans (18–24 years old) use other platforms like Instagram, Snapchat and TikTok more frequently than sites like Facebook (Auxier & Anderson, 2021). Americans' use of social media has grown from just 5% in 2005 to 72% today (Pew Research Center, 2021). Previous research on social media use and NSSI behavior has been limited to exploring the benefits and the risks of the usage of social media for NSSI. Much of the prior research has been very similar, in that it has focused on investigating the benefits and risks of posting about NSSI behaviors and pro-NSSI Instagram pages. Some studies have taken a balanced approach; on the one hand, they list the benefits of reducing social isolation and finding social support, while on other hand they also find that posting NSSI activity can invite reinforcement of NSSI or triggering NSSI urges (Lewis & Seko, 2015).

A review of the literature on both non-suicidal self-injury and social media use revealed some key points to consider. These points are: (1) definition of non-lethal self-injury, (2) prevalence of pro-NSSI posts on social media, (3) pervasiveness of social media impacts people's behavior and choices, (4) mental health and the impact of the internet and (5) exploring use of LIWC as new method of understanding language online. Finally, the research questions will be detailed.

There are varying definitions of self-harm in the literature. One of the most widely cited resources for self-harm/self-mutilation is by Favazza (1998) which according to Google Scholar has been cited 1269 times in the professional literature. Favazza defines self-mutilation as "the deliberate, non-suicidal destruction of one's own body tissue" (p. 260). He concludes his overview by writing, "Superficial/moderate SM includes compulsive acts such as trichotillomania and skin picking and such episodic acts as skin-cutting and burning" (p. 259). As social media rises in popularity and accessibility, there has been a corresponding increase in social media posts about NSSI. Adolescents and emerging adults have been posting pictures of their NSSI on social media and posting text that describes NSSI behavior, including tips on what to use to self-injure and how to hide their injuries. In a meta-analysis exploring the risks and benefits of online posting of NSSI, Lewis and Seko (2015) found major themes in NSSI social media involvement. Their study found that one possible reason that people who utilize NSSI post on social media is to connect with others who are going through similar experiences. Many people who engage in NSSI experience high levels of shame, fear and embarrassment. One argument from the Lewis and Seko study is that people who utilize NSSI behaviors can post anonymously without judgment and shame. The researchers also found that there was a "reduction of social isolation through online interactions" (Lewis & Seko, 2015, p. 254). While prior studies have attempted to understand the reasons and consequences for discussing NSSI on social media, there has been little or no engagement about the language that they use to describe their NSSI and what that language can tell us.

The pervasive nature of social media provides a normative influence, affecting people's social behaviors, consumer habits, personal choices and even mood (Hartmann, 2019). In 1955, Deutsch and Gerard described the phenomenon known as "normative influence" as an intrinsic characteristic of human behavior, in which social influences lead to conformity. While much has changed in American society since 1955, the concept of conformity is very much alive in the technological age. Huh et al. (2014) have further developed the concept of normative influence by describing a phenomenon known as the "social default", which includes how other people's known decisions will impact a person's personal choices. These concepts are grounded in people's natural tendency to mimic other people's behaviors and describe how observing other's choices can create social default effects. Businesses in the age of widespread social media usage, seeking to harness this tendency to mimic others for their own bottom lines, have studied these human behaviors to understand the effects of social media on consumer choices; they have used the results of their studies to inform and advance their marketing strategies. Mental health, however, has lagged behind business, and appears to continue to lack in understanding just how social media influences mental health behaviors. Lavis and Winter (2020) disagrees with this view point. In their study, they explored whether social media exposure leads to self-harming behavior or if self-harming behavior leads to online social media engagement. Lavis and Winter discovered in their data that self-harming behavior is often used prior to social media engagement in which the authors of social-media posts were in an effort to seek peer support and participate in a self-harming sub-culture. Regardless of the influence direction of social media, it is evident that social media influences mental health either positively or negatively.

Online behaviors seem to both fascinate and astound social media consumers. We can see how observing other's choices being documented on social media affects individual's decision-making processes. While many are aware of social media challenges, little is known about why or how these challenges become popular (Khasawneh et al., 2021). Social media challenges range from harmless to helpful. In opinion pieces and elsewhere, people have speculated that peer pressure and pressure to increase one's social media popularity and gain followers have driven the widespread, dangerous popularity of these challenges (Setzekorn, 2018). While there is no definitive answer on why these deadly behaviors become glamorized on social media, it is clear that they reveal that there is an online pressure to conform and that people can be influenced even by online evidence of other's decisions when making their own personal choices.

Social media can also allow people to become involved in online communities, like Reddit. In the first half of 2021 Reddit recorded 1.7 billion visits and host over 130,000 digital communities (Clement, 2021). Shanahan et al.'s (2019) study on images of self-harm on social media found that many who use self-harm use social media to express their negative feelings and experiences. Unlike blogging platforms like Tumblr, the website provides its users with anonymity, which may make them feel more protected; the only requirement for posting on Reddit is a screen name (Harriagian, 2018). Each community in Reddit, which is called a subreddit, allows people to post on a topic, and others can respond to these posts. Reddit's anonymity may encourage people to post more honestly about their experiences and feelings. Reddit's anonymity and its users' increased comfort level makes Reddit a good fit for a linguistic analysis study (Kamarudin et al., 2018). For the purpose of this study, the subreddits explored and used in the corpus involved pro-NSSI behavior.

As social media usage rises, mental health researchers have become more interested in how people's language use can give indications of their mental health. In a study on social media posting by people struggling with post-partum depression, Choudhury et al. (2013) noted that the language that these participants used provided helpful "psychological markers" that gives information on the individual's inner workings.

Similarly, there has been an increased effort among mental health experts to understand how the language of self-harm has developed and is expressed on social media.

This study uses the corpus linguistics method to evaluate the use of NSSI related language on Reddit. This approach uses a corpus constructed from large amounts of data from any text source, like social media posts. The data is then inputted into software for analysis; the analytical software can complete a more in-depth analysis than what people are capable of doing (McGlashan, 2018). The Linguistic Inquiry and Word Count (LIWC; Pennebaker et al., 2007) is commonly used in current literature because it is capable of analyzing a variety of categories and emotional words; it is also capable of analyzing and calculating aspects of language. LIWC is often used to explore mental health markers in which researchers specifically focus on emotion word use, social word use, self-referencing, drives and pronoun use (Adrian et al., 2011). Analyzing language has the potential to inform research into an individual's experience, environment and mental health.

Given the aforementioned, three research questions were designed to guide this study. These questions were:

**RQ1**: What is the NSSI specific content word use of the individuals making posts?

**RQ2**: Does use rate of linguistics processes in NSSI posts differ from the general use rate of these processes in blogs?

**RQ3**: Does use rate of psychological processes in NSSI posts differ from the general use rate of these processes in blogs?

## Method

### Design

This study used synchronic corpus linguistic design to explore public Reddit posts on NSSI (Weisser, 2016). There were 16 variables that fell across 3 categories (NSSI-specific, linguistic processes psychological processes. The level of measure for all variables was continuous. The unit of analysis was individual words. Because the study used publicly accessible texts and individual identifier were not web scraped, this research did not require human subject's review.

### Study Corpus

This corpus was constructed using Reddit's API system to collect public posts that were published between January 1, 2017 and December 31, 2017. This API was instructed to scrape posts that: (a) were written in monolingual English and (b) included words or linguistic features from the list of NSSI terms compiled by the researchers. The API was instructed to eliminate user names, URLs, hashtags, location of the posts, posts from outside of the United States, photographs and foreign languages while collecting texts for the corpus. The number of Reddit posts collected was 1,322. A number of preprocessing steps (e.g., abbreviations were spelled out) took place to prepare this corpus for analysis and a list of these steps can be reviewed on this study's Open Science Foundation research project webpage (https://osf.io/xxxxx/). Because the definition of what is a word differs among linguistic software packages, the total word count for the corpus slightly varies between RQ1 ($n$ = 227,946) and RQs 2-3 ($n$ = 230,161). Since the data is reported in normalized frequencies this difference is inconsequential when interpreting the results.

### Reference Corpus

The Reddit was compared to the national norms for blog posts recorded in Pennebaker, Boyd et al. (2015). The counts were not whole numbers because means for the posts were reported. No preprocessing was required.

### Measures

#### ***XXXX (blinded for review) NSSI Term Dictionary (XNTD)***

The level of use of NSSI-specific terms was assessed by means of the XXXX NSSI Term Dictionary (XNTD). This dictionary has six sub-dictionaries: Methods Except Cutting (e.g., burning), Cutting Methods (e.g., cut, carve), general NSSI terms (e.g., NSSI), Instruments (e.g., blade), Reasons (e.g., anger), and Scar (e.g., scarification). On the research project's webpage (https://osf.io/xxxxx/) two supplemental files concerning this dictionary can be view: (a) a list of all dictionary words and word stems in Regex format, and (b) a list of words and word stems by each subdictionary. The word counts were accomplished using #Lancsbox software program (Brezina et al., 2020). The resultant list was reviewed for words produced by stemming but irrelevant to the study (e.g., scar* producing scary). The irrelevant words were eliminated prior to analysis.

#### ***Linguistic Inquiry and Word Count (LIWC)***

Linguistic and psychological processes were assessed by means of the LIWC software program (Pennebaker, Booth et al., 2015). The linguistic processed assessed included the following pronouns: 1st person singular, 3rd person singular, 1st person plural pronoun, and 3rd person plural. The psychological processes included the following pronouns: Negative Emotion (e.g., hurt), Anxiety (e.g., worried), Anger (e.g., hate), Sadness (e.g., crying), Risk (e.g., danger). Pennebaker, Boyd et al. (2015) reported adequate reliability and validity for all LIWC subscales.

### Data Analysis

In terms of RQ1 (NSSI-specific variables), both raw counts and normalized counts (percentage of all words) were reported.  Regarding RQ2 (linguistic variables) and RQ3 (psychological variables), the study corpus will be compared to the general blog corpus by means of the log likelihood ratio test ($G^2$). This test requires raw scores which were obtained by multiplying the percentage of all words and corpus size given in the LIWC output. The effect size measure was log ratio (LR). As Hardie (2014) notes, a log ratio of 1 means a word is twice as common in a corpus than it is in the comparison corpus. All analyses were conducted using Excel with a standard .05 alpha.

## Results

Regarding RQ1, Reason was the highest frequency category (1.57%), followed by Cutting-Specific Terms (0.68%). Complete results can be inspected in Table 1. The raw counts for each word and word stem used in the XNTD can viewed on the project's research webpage (https://osf.io/xxxxx/). Concerning RQs 2-3, the linguistic and psychological processes that most differed from the general blog norms were first person singular ($G^2$ = 68.19, LR = .77) and negative emotion ($G^2$ = 42.00, LR = 1.02). Complete results are also listed in Table 1.

## Discussion

This study examined language used in pro-NSSI micro-blogs on Reddit and investigated whether there were differences between the use of language in pro-NSSI posts when compared to blogs overall. For each research question, probable reasons for the obtained results are presented and discussed. Finally, limitations and implications for menta health professionals are presented.

In terms of RQ1, the findings indicated that the words most frequently used in pro-NSSI posts were reasons for their NSSI. In this category, the most frequently used words in rank order are feel and help. In the data drawn from posts, it seems as though the users share a strong desire to get help for their NSSI behaviors, but there is also a high focus on their scars. Frequently, it is very difficult for people engaged in NSSI to talk about their struggles with their close friends and family; they often experience a lot of behaviors from others, such as questioning why the individual can't stop self-harming or expecting them to stop the self-harming behaviors quickly, that add to the internal shame they already feel (Kane, 2017). Many of the blogs discuss how outside people react negatively to NSSI behaviors, which triggers an intense emotional reaction. It appears that Reddit provides a safe place to share about one's experience without the fear of judgement and shame.

Another possible reason for the high frequency of the words feel and help may be the individual's intentions regarding self-injury. One of the long-standing myths about NSSI, which is still held by many mental health professionals, is that NSSI is related to suicidal intent (Klonsky et al., 2014). While the word suicide is mentioned 86 times in the raw data, an analysis of the context reveals that suicide is mostly used when individuals are expressing that they do not want to attempt suicide, they want to end the pain they are experiencing now. The data suggests that many people who are engaged in NSSI feel misunderstood about their behavior and insist that they are not suicidal. This study supports the findings in previous research that adolescents who engage in NSSI communicate more about their behavior with friends and mental health professionals than adolescents who engage in suicidal self-injury (Baetens et al., 2011). These findings suggest that while the individuals are not seeking to end their lives, they are seeking help. While both the anonymity theory and the intention theory attempt to explain the results of specific NSSI word use, it is more likely that the language is can be attributed to the intention theory than that anonymity theory.

The method of self-harm is widely contested in the literature. Chartrand et al., (2016) found that there was no statistical difference in the use of cutting to self-harm versus self-poisoning. However, Shanahan et al., (2019) discovered in their analysis of social media images that cutting appears to be the preferred method of self-harming behavior. While it is unclear of the motivation to cut to self-harm or another method of self-harming behavior, it is clear that the most predominate focus of self-harm on social media is that of visible scars made by self-inflicted cuts (Shanahan et al., 2019).

The raw data from Reddit reveals that individuals making pro-NSSI posts use language that is intensely focused on their scars, though some users talked about their scars in a prideful manner and others talked about their scars in a shame-filled, negative manner. One possible reason for this is the anonymity of Reddit; it is likely much easier for a person who uses NSSI to talk about their scars anonymously online rather than talking about them in person.

RQ2 concerns the linguistic patterns in pro-NSSI blogs. The data demonstrates that the most frequent and statistically significant pattern was the use of the first-person singular—words such as I, me and my. According to Pennebaker (2013), people's pronoun use tracks their focus of attention. Therefore, people who use first person singular pronouns are typically looking at themselves quite often and most likely experience anxiety, self-consciousness, pain and depression. People who engage in NSSI usually suffer from a form of mental illness like anxiety or depression and use NSSI to cope (Skegg, 2005). It stands to reason that people who use NSSI to cope also struggle with mental health issues, and this is reflected in their pronoun use.

Another reason that Reddit users discussing NSSI may use the first-person singular is the social isolation of NSSI. Many individuals who engage in NSSI self-harm in private and often feel a lot of shame about their behavior. The first-person singular pronouns demonstrate a lack of community or sense of belonging; instead, they express a feeling of aloneness. Pennebaker (2013) also stated that "I" words denote vulnerability. Reddit is a unique platform that allows its users to post on their website with a simple, fictional user name. It is one of only a few anonymous message board forums. This gives their users a sense of security because it would be difficult to identify who they are simply using a screen name. The two reasons support each other in the results, but it is likely that the most frequent use of first-person singular pronouns is due to the isolation these individuals experience.

RQ3 pertains to the psychological word use patterns in pro-NSSI blogs. The most common and statistically significant psychological pattern was negative emotion language. Pennebaker (2013) suggests that through his findings in research younger people tend to use more negative emotion words when they write, unlike older people who write using more positive emotion words. He discovered that this difference in word usage between older and younger writers became apparent around the age of 40, and then became much more pronounced in the oldest age groups. Pennebaker theorized that older writers can better regulate their emotions and can look at the world from different perspectives. Given this research, it stands to reason that the people who post on pro-NSSI Reddit blogs are younger, more impulsive and view the world through a darker lens.

Another possible reason why those engaged in NSSI use a lot of negative emotion words is high levels of neuroticism. Pennebaker et al. (2003), discovered that neuroticism was positively correlated with the high use of negative emotion words. This finding was also consistent with the high use of first-person pronouns and the amount of text in which the individual is self-focused. Neuroticism is one of the big five personality traits (Costa & McCrae, 1985). Using Eysenck's influential Psychoticism, Extraversion and Neuroticism theory of personality (Eysenck, & Eysenck, 1985), Shatz (2004) noted that neuroticism involves differences in emotional reactivity to negative environmental stimuli with high scorers exhibiting anxiety, depression, guilt, low self-esteem, and moodiness. The research from Pennebaker et al. supports this study's findings that demonstrate that people who are engaged in NSSI use more negative emotion words and first-person pronouns. While it is likely that both age and neuroticism cause this high usage rate of negative emotion words are very likely, but neuroticism is the more likely cause.

The study's results produce an image of a typical NSSI-engaged person as a young individual who is affected by a form of mental illness, experiencing feelings of isolation and attempting to cope with their intense feelings. This study, like others, found that the posts written by individuals who engage in NSSI contain psychological markers, namely the use of negative emotion language, that demonstrate a level of mental illness. This negative emotion language also demonstrates that the user is experiencing a sense of longing for a safe place to talk about topics that may not be socially acceptable, a sense of aloneness and isolation and an indication of age and perspective. Therefore, this research implies that using the client's strengths which is written expression, and reframing in more positive terms can benefit the client's mental health and adaptive coping behaviors.

There are three limitations to the study that should be noted. First, words often have multiple meanings and could be put into a category that does not reflect the true meaning of the word in its context. Second, the amount of data that was included in the corpus could have been limited by the fact that only English language words were included. Third, this study may not be generalizable to the entire population of people who post on pro-NSSI social media and blog sites. Inferences into the data should be done with care.

The results contain X implications for mental health professionals. First, there are several persistent myths about NSSI in the mental health community. One of those myths is that people who use NSSI to cope are also suicidal. This study reveals that while NSSI engaged persons may experience suicidal thoughts, they are not trying to end their lives. These results support previous research that has also distinguished between NSSI and suicidal attempts, though this myth endures in the field of mental health (Brown et al., 2013; Wood, 2009). When posts on Reddit are evaluated linguistically, it seems clear that people who engage in NSSI are directly communicating with a broader community and saying that they are not suicidal and that they are not trying to end their life. Counselors may be better able to understand the true intentions of people who are engaged in NSSI, and thus improve their treatment plans, by examining the actual words used by these individuals.

Analyzing and understanding the language used by individuals to describe their mental health and NSSI behaviors can help mental health professionals in the mental health field better understand their patients and provide insight into their inner world. This study's results demonstrate how we as mental health professionals can better communicate with these individuals and treat them. While research (Skegg, 2005; Shanahan et al., 2019; Wood, 2009) and lived experience have already demonstrated that people who engage in NSSI are usually younger, we now can identify their language markers to be more negative in tone and see that they use more first-person singular pronouns which gives mental health professionals insight into how alone they feel. This has the potential to help mental health professionals to explore better and more efficiently the level of isolation, aloneness and level of mental illness, improving treatment.

This study found that people who engage in NSSI and post about it on blogs are seeking help by reaching out to a community of similar people. This study implies that these people are already using writing as one method for coping with their emotional pain. Pennebaker (2013) found in his study that people who used positive emotion words while writing about their personal trauma experienced improved physical and mental health during the weeks of the experiment. In this same study, researchers discovered that people who used negative emotion words frequently did not benefit from writing.

Mental health professionals can use the information from this study to inform their practice by making use of theories that may better reach persons who use NSSI. There is a high number of adolescents and young adults who post their NSSI behaviors on blogs. If we are to meet them where they are at, it is on the internet writing about their pain. However, their writing is filled with negative emotion words, which Pennebaker (2013) suggests may prevent this coping mechanism from being useful. One possible way that mental health professionals can use the insights from this study is by using a theory such as Narrative therapy to utilize what clients are already doing and altering their writing to be more beneficial to them. Hoffman and Kress (2008) have suggested that Narrative therapy is a beneficial treatment for those who use NSSI because this approach gives clients the power to name the problem and helps them learn to separate the problem from themselves, and this type of intervention can effectively be used with this population.

**Table 1**

*Results from RQs 1-3*

| Category | Process Type | Reddit Word Ct. | Percent of A.W. | Blog Norm Word Ct. | Over/ Under Use | $G^2$ | LR |
|---|---|---|---|---|---|---|---|
| 1st P. Sing. | Ling. | 24581.19 | 10.68 | 200.72 | + | 68.19 | 0.77 |
| Neg. Emot. | Psy. | 9620.73 | 4.18 | 66.05 | + | 42.00 | 1.02 |
| 1st P. Pl. | Ling. | 782.55 | 0.34 | 29.18 | - | 20.48 | -1.42 |
| Anxiety | Psy. | 1680.18 | 0.73 | 8.66 | + | 12.15 | 1.43 |
| Sad | Psy. | 2117.48 | 0.88 | 14.11 | + | 9.86 | 1.06 |
| Risk | Psy. | 2140.50 | 0.92 | 14.75 | + | 9.27 | 1.02 |
| Anger | Psy. | 2025.42 | 0.93 | 21.80 | + | 1.56 | 0.37 |
| 3rd P. Pl. | Ling. | 1427.00 | 0.62 | 21.80 | - | 0.18 | -0.13 |
| 3rd P. Sing. | Ling. | 3383.37 | 1.47 | 48.10 | - | 0.02 | -0.03 |
| Meth. Ex C | NSSI | 828.00 | 0.0036 | | | | |
| Cutting | NSSI | 1548.00 | 0.0068 | | | | |
| Terms | NSSI | 598.00 | 0.0026 | | | | |
| Instruments | NSSI | 143.00 | 0.0006 | | | | |
| Reasons | NSSI | 3572.00 | 0.0157 | | | | |
| Scar | NSSI | 460.00 | 0.0020 | | | | |

*Note.* Corpus sizes for inferential analysis: Study $n$ = 230,161; Reference $n$ = 3206.45. Critical value for $G^2$ at $p < .05$ is 3.84. Corpus size for NSSI descriptive statistics for was $n$ = 227,946.